\documentclass[article,showpacs]{revtex4}

\usepackage{setspace}
\usepackage{times}
\usepackage{graphicx,colordvi}

\def\gtorder{\mathrel{\raise.3ex\hbox{$>$}\mkern-14mu
 \lower0.6ex\hbox{$\sim$}}}
\def\ltorder{\mathrel{\raise.3ex\hbox{$<$}\mkern-14mu
 \lower0.6ex\hbox{$\sim$}}}

\def\ge{G_E}
\def\gm{G_M}

\def\gegm{G_E / G_M}

\begin{document}

\title{Evaluation of the proton charge radius from e-p scattering}



\author{John Arrington}
\affiliation{Physics Division, Argonne National Laboratory, Argonne, IL 60439}
\author{Ingo Sick}
\affiliation{Dept. f\"ur Physik, Universit\"at Basel, CH4056 Basel, Switzerland}

\begin{abstract}

In light of the proton radius puzzle, the discrepancy between measurements of
the proton charge radius from muonic hydrogen and those from electronic
hydrogen and electron-proton scattering measurements, we reexamine the
charge radius extractions from electron scattering measurements. We provide a
recommended value for the proton RMS charge radius, $r_E = 0.879 \pm
0.011$~fm, based on a global examination of elastic e-p scattering data. The
uncertainties include contributions to account for tension between different
data sets and inconsistencies between radii using different extraction
procedures.

\end{abstract}

\maketitle


\section{Introduction}

Multiple extractions of the proton RMS charge radius have been made based
on global analyses of world data on elastic e-p scattering. The radius
is related to the slope of the charge form factor, $\ge(Q^2)$, at $Q^2=0$.
As such, the extraction is sensitive to the precision of the data available
at low $Q^2$, the minimum $Q^2$ value of the data set, the uncertainty in the
absolute normalization of the data, and the model dependence of the
fitting function used to perform the extrapolation to $Q^2=0$. Many radius
extractions have not examined these questions in detail or have not included
uncertainties associated with the model dependence of the extraction, yielding
a significant range of quoted values for the charge radius, often with
unrealistic uncertainties. More recently, a high-precision data set from Mainz
has significantly expanded the body of data at low $Q^2$~\cite{bernauer10}.

Extractions which attempt to account for all of these uncertainties appear in
Refs.~\cite{sick03, sick06, zhan11, ron11, sick12, bernauer13}.  All of these
extractions yield proton RMS charge radius values of approximately 0.88~fm,
consistent with those extracted from atomic physics measurements in electronic
hydrogen~\cite{mohr12}. The 2012 CODATA evaluation~\cite{mohr12} combining e-p
scattering and atomic transitions in hydrogen yields $r_E=0.8775(51)$~fm, 4\%
higher than the extremely precise muonic hydrogen result~\cite{antognini13}
which yields $r_E=0.84087(39)$~fm, and inconsistent at the 7$\sigma$ level.
The exact results from several of these extractions, along with our updated
recommendation, are given in Table~\ref{tab:radii} in Sec.~\ref{sec:summary}.

Significant work has gone into attempting to understand the proton radius
puzzle, examining corrections to atomic transitions in hydrogen and
muonic hydrogen, as well as proposing explanations in terms of new physics.
For detailed reviews, see Ref.~\cite{pohl13, carlson15}. In this work, we
examine recent extractions of the proton radius from e-p scattering data and
suggest a combined result and uncertainty that accounts for our current
understanding of the systematic uncertainties and model dependence in such
extractions. This will allow for a more reliable comparison of the scattering
results, the atomic hydrogen measurements, and the muonic hydrogen results.

\section{Extraction of the radius from e-p scattering}

Elastic e-p scattering depends on the proton charge and magnetic form factors
$G_E(Q^2)$ and $G_M(Q^2)$ which, in the non-relativistic approximation, are
the Fourier transforms of the charge and magnetization densities. In this
approximation, the RMS radius of the charge density is related to the slope of
the charge form factor at $Q^2=0$. Relativistic effects yield model-dependent
corrections which invalidate this simple connection between the charge density
and form factors. For the extractions examined here, the conventional approach
was adopted, where the radius is defined in terms of the low-$Q^2$ expansion of
the form factors as
\begin{equation}
\ge(Q^2) = 1 - \frac{1}{6}r_E^2 Q^2 + ...~,
\label{eq:definition}
\end{equation}
allowing measurements of the form factors at low $Q^2$ to be used to extract
the radius. In the one-photon exchange approximation, the spin-independent
cross section depends on $G_E^2$ and $G_M^2$, with the angular dependence at
fixed $Q^2$ allowing for separation of $\ge$ and $\gm$. In cases where one
form factor dominates the cross section, spin-dependent cross section data,
which are sensitive only to the ratio $\gegm$, provide improved extraction of
the form factor with smaller contributions to the spin-independent
scattering~\cite{arrington07a, perdrisat07, arrington11a}

Most elastic e-p experiments provide a set of cross sections and uncertainties
along with an estimate of the overall normalization uncertainty which applies
to the entire data set. A global analysis will often allow each of these
normalization factors to vary, with a contribution to the fit's total $\chi^2$
associated with the size of the normalization adjustment. The recent Mainz
data~\cite{bernauer13} has 31 normalization factors which are are
completely unconstrained and must be determined in the fit, with the
extrapolation to the known value $\ge(0)=1$ providing the absolute
normalization. 

Within the world data set, excluding the recent Mainz
measurement~\cite{bernauer13}, the individual experiments show consistency
within the quoted errors~\cite{arrington03,
sick03}. However, a comparison of the Mainz form factors with world data
clearly indicates that the Mainz data are not entirely compatible with the
previous measurements, complicating the task of presenting a combined result.
In addition, the way in which the Mainz data set quotes uncertainties is
different from essentially all previous measurements. Simply treating the
quoted uncertainties on the Mainz and world cross section measurements
on equal footing would strongly bias the result in favor of the Mainz data,
as discussed in more detail below. Because of this, we will examine the radius
extraction from the Mainz data set and other world data separately, and then
compare the results to examine the question of consistency.

\subsection{Mainz data set}

First, we note that because of the large number of data points in the Mainz
data set~\cite{bernauer13}, coupled with the lack of measured absolute
normalization of the cross sections, the A1 collaboration presented the data
as several data subgroups, each with its own normalization, and separately
estimated uncorrelated and correlated systematic uncertainties. The
uncorrelated systematic uncertainties account for possible errors that are
independent for every cross section and thus are treated identically to the
statistical uncertainties. The correlated systematic uncertainties account for
potential corrections which have a correlated effect across many measurements
(but excluding pure normalization uncertainties). They chose to model the
correlated systematics by
applying corrections which depend linearly on the scattering angle, and which
typically vary by 0.2\% between the largest and smallest angle of each data
subgroup.

The separation of the uncertainties into uncorrelated, correlated, and multiple
normalization uncertainties makes the extraction of the radius from the data
rather complex. The data as provided in Ref.~\cite{bernauer13} already applies
normalization factors determined in the global fit to the Mainz data, and 
no information on the original normalization is available. Therefore, any
analysis of these data requires that the normalization factors of each subset
be allowed to vary as part of the fit, as well as separate treatment of the
uncorrelated and correlated systematic uncertainties. Using fixed
normalization factors rather than allowing them to float as part of the fit is
not consistent with the way the uncertainties are estimated by the experiment
and can artificially decrease uncertainties by a factor of
5 or more~\cite{lee15}.

Several values of the charge radius have been extracted based on the Mainz
data~\cite{lorenz12, bernauer13, lorenz14, carlson15, lorenz15}, and these
analyses yield a significant range of results for $r_E$ depending on the
treatment of systematic uncertainties and normalization factors, choice of
two-photon exchange corrections, and choice of fitting function used. However,
most of these do not fully treat the uncertainties at the level required to
obtain a reliable estimate of the uncertainty. The only complete treatments
published so far are those of refs.~\cite{bernauer13, lorenz15}, though
another complete analysis which also includes a detailed examination of the
systematic uncertainties has recently been completed~\cite{lee15}.
Ref.~\cite{lorenz15}
yields very different results for model-independent fits using the
$z$-expansion~\cite{hill10} and a dispersion relation analysis, and while their
dispersion relation approach yields a radius that is consistent with the muonic
hydrogen radius, it provides a much poorer fit to the Mainz data. Because of
this, we will consider only the radius extractions shown in
Ref.~\cite{bernauer13} in our examination of the Mainz data, although we will
examine the assumptions and error treatment used in these extractions.

As noted above, the Mainz breakdown of the uncertainties makes it difficult to
perform a combined analysis of Mainz cross sections and other world data. In
such fits, the relative weighting of the different measurements is determined
by the uncertainties applied to each data point. For the Mainz data, this is
only the statistical and uncorrelated systematic uncertainty, which represents
a small fraction of the total uncertainty, while other experiments include
everything except the normalization uncertainty in these errors, including
those associated with radiative corrections and two-photon exchange, neglected
in Ref.~\cite{bernauer13}. Thus, in a naive global cross section analysis, the
Mainz data will receive far more weight than it should compared to all other
experiments.

\subsection{World data}

Data from several scattering experiments are available, e.g. see
Refs.~\cite{arrington03, arrington04a, arrington07a, perdrisat07, arrington07c,
arrington11a}, and there have been many parameterizations made of the proton
form factors based on these data. In several cases, the goal was to have a
global parameterization of the full data set, up to $Q^2 \approx 30$~GeV$^2$
for $\gm$. Commonly used global parameterizations~\cite{budd03, kelly04,
bradford06, arrington07c} did not attempt to ensure that the fit function was
flexible enough to provide a reliable extraction of the radius in the presence
of a large body of higher-$Q^2$ data. One exception is
Ref.~\cite{venkat11} which performed a low $Q^2$ fit to constrain the slope,
and then performed the global fit with the slope parameters fixed. Because we
are not interested in the form factor at high $Q^2$, we will examine only fits with the
explicit aim of extracting the proton radius from low-$Q^2$ data.

In determining the radius from world data, we rely on the results from two of
the more recent extractions~\cite{zhan11, sick12} which focus on the low-$Q^2$
region, examine the model dependence of the extraction, and use a relatively
complete data set including polarization data. There are several other
analyses which could be included but which we do not use here.
Single-parameter fits are excluded as the $Q^2$ range necessary to provide a
useful limit on the radius goes beyond the region where the single parameter
fits are sufficiently precise to represent the data~\cite{sick03, kraus14}.
Some fits~\cite{sick03, blunden05b, sick06, arrington07b} are excluded because
they are essentially earlier versions of the fits which we do include, and as
such do not add much new information. Others are excluded because they include
only a limited set of the low $Q^2$ data, they apply two-photon exchange
corrections based on phenomenological extractions which are not well
constrained at low $Q^2$~\cite{graczyk14}, or because they fit to extracted
form factors rather than the original cross section measurements~\cite{hill10,
epstein14} which generally means neglecting normalization uncertainties on
the different data sets.

\subsection{Two-photon exchange corrections}

One critical issue is the treatment
of two-photon exchange corrections~\cite{carlson07, arrington11b} in the
extraction of the charge radius~\cite{blunden05b, arrington11c, arrington13}.
The Mainz extraction~\cite{bernauer13} applies the correction of
Ref~\cite{mckinley48}, which is valid (up to corrections of order
($Z\alpha)^2$) only for scattering from a point nucleus.  For the proton, the
higher order corrections are negligible and in the limit $Q^2 \to 0$, the
scattering is not sensitive to the internal structure, and thus the point-proton
approximation is appropriate. At non-zero $Q^2$ values, the correction changes
in both the Coulomb distortion correction for finite-size nuclei~\cite{Sick98}
and the extension of hard TPE corrections to a proton with internal degrees of
freedom, yielding a $Q^2$-dependence~\cite{blunden05a, carlson07,
arrington11b, arrington13} which is clearly relevant in attempting to
determine the slope of $G_E(Q^2)$ as $Q^2 \to 0$.

Note that the correction at $Q^2=0$ has the opposite sign and angular
dependence compared to what is necessary to explain the discrepancy between
Rosenbluth and polarization form factor measurements at larger $Q^2$
values~\cite{guichon03, arrington03, arrington04a, perdrisat07, arrington07a,
arrington11b}. Calculations at finite $Q^2$ which account for the structure of
the proton show this change of sign in going from $Q^2=0$ to larger $Q^2$
values~\cite{blunden05a, kondratyuk05, kondratyuk07, zhou14}, but there is a
significant model dependence associated with the modeling of the intermediate
hadronic state. Recent comparisons of electron-proton and positron-proton
scattering confirm a change of sign by $Q^2=1$~GeV$^2$~\cite{adikaram15,
rachek15} and are in reasonably good agreement with the TPE calculations
performed in a hadronic basis~\cite{blunden05a, kondratyuk05, kondratyuk07,
zhou14}. In addition, all TPE calculations which are expected to be valid at
low $Q^2$ values are in very good agreement below
$Q^2=0.2$~GeV$^2$~\cite{arrington13}. Given the consistency of the
calculations and the experimental demonstration that there is a significant
$Q^2$ dependence in the TPE effects at low $Q^2$, we will use only extractions
which include hadronic TPE corrections, and include an estimate for the model
dependence of these corrections.

\section{Examination of previous charge radius extractions}

In the following sections, we examine the existing extractions of the proton
radius presented above and provide recommendations for updated radius
values from Mainz and world data based on the inclusion of additional
uncertainties and corrections. We also provide a simple combined result as
well as a suggested combined result with additional uncertainty intended to
account for the observe discrepancies between the Mainz and world data sets.

\subsection{Radius extraction from Mainz data}

We chose to take the extracted values of charge and magnetic radii from the
analysis of Ref.~\cite{bernauer13} which includes the hadronic TPE corrections
of Ref.~\cite{blunden05a}:
\begin{equation}
r_E = 0.875(5)_{\rm stat}(4)_{\rm syst}(2)_{\rm model}(5)_{\rm group} ~, 
r_M = 0.799(13)_{\rm stat}(9)_{\rm syst}(5)_{\rm model}(3)_{\rm group} ~,
\end{equation}
where the breakdown of the uncertainty is detailed in Ref.~\cite{bernauer13}.
Note that it is not clear in~\cite{bernauer13} how these uncertainties should
be combined. While most authors combine these contributions in quadrature,
yielding $\delta r_E = 0.008$~fm, we note that all figures in
Ref.~\cite{bernauer13} that show the combined uncertainties on the radii or
form factors combine the errors linearly, corresponding to $\delta
r_E=0.016$~fm. When asked about this, the response from the Mainz A1
collaboration was that the first three uncertainties should be combined in
quadrature while the final contribution should be added
linearly~\cite{priv_distler}, yielding $r_E=0.875(12)$~fm, $r_M=0.799(20)$~fm.

As noted earlier, the quoted radius includes no uncertainty associated with the
TPE correction. The impact of replacing the McKinley-Feshbach
correction~\cite{mckinley48} with the hadronic calculation~\cite{blunden05a}
was small, decreasing $r_E$ by 0.004~fm and increasing $r_M$ by 0.022~fm. We
take half of this shift as an additional uncertainty to account for the model
dependence of the TPE corrections.

Refs.~\cite{arrington_theseproc, lee15} demonstrate that impact of correlated
systematic errors on the radius can be significantly larger if the correction
is assumed to have a different functional form different from that used
in~\cite{bernauer13} (linear with $\theta$) or if it was applied to different
subsets of the data (e.g. to each of the 34 data subgroups with different
normalization factors rather than each of the 18 spectrometer-beam energy
combinations). Based on these tests, and concerns about other sources of
correlated systematic uncertainties~\cite{sick12, arrington_theseproc, lee15}
that are neglected in the extraction, we double the systematic uncertainty
from on $r_E$ from 0.004 to 0.008~fm, and on $r_M$ from 0.009 to 0.018~fm.

With the additional TPE uncertainty, the increased correlated systematic
applied to both charge and magnetic radii, and using the proposed procedure
for combining the different sources of uncertainty, we obtain the following
radii from the Mainz data: $r_E = 0.875(15)$~fm, $r_M = 0.799(28)$~fm.

\subsection{Radius extraction from world data}

For radii extracted from world data, excluding the recent Mainz cross sections, we
take the extractions from Ref.~\cite{zhan11}, $r_E=0.875(10)$~fm, and
Ref.~\cite{sick12}, $r_E=0.886(8)$~fm. Note that the result quoted in
Ref.~\cite{sick12} was from a combined analysis of Mainz and world data, with
additional systematic uncertainties applied to the Mainz data to account for
the uncertainty in the subtraction of the target window contribution. The
combined analysis yielded a radius that was larger by 0.001~fm than the result
quoted here which excludes the Mainz data.

These analyses include very similar data sets,
with differences mainly associated with the choice of fitting function and the
evaluation of the model dependence. We choose to take an unweighted
average of the two results, $r_E=0.881(9)$, and apply an additional
uncertainty of 0.006~fm, equal to half of the difference between the two
results, as an estimate of possible model dependence beyond what is included
in the analyses. For the magnetic radius, the value of~\cite{zhan11} is taken,
as Ref.~\cite{sick12} did not provide a value of $r_M$ while the previous
work~\cite{sick06} did not include the more recent polarization measurements
which help constrain $\gm$ at low $Q^2$. This approach described above yields
a final world data radii of $r_E=0.881(11)$~fm and $r_M=0.867(20)$.

\subsection{Combined e-p scattering radius}

Simply combining the charge radii from Mainz and world data yields
$r_E=0.879(9)$~fm. However, while the results for the
charge radius from the Mainz and world data sets are consistent, there are
significant tensions between the Mainz and world data, e.g. Figs.~10 and 19 of
Ref.~\cite{bernauer13}. In addition, the Mainz and world updated magnetic
radii differ by three standard deviations as quoted in Refs.~\cite{zhan11,
bernauer13}, and two standard deviations in our updated result (due to a
combination of a slight reduction in the inconsistency and an increase in the
quoted uncertainties). Given the tension between the results, it is not clear
how one should treat the uncertainty on the combined result, and we are forced
to make a relatively arbitrary decision on how to do so.

We choose take the discrepancy in $r_M$ and use this to estimate the potential
systematic errors in $r_E$, based on the relative sensitivity of $r_E$ and
$r_M$ to changes in the data. Examining the variation of the radii when
applying corrections (e.g. TPE contributions) or varying fit functions, we
find that the magnetic radius has a sensitivity that is roughly 10 times
larger than that of the charge radius. We thus take 10\% of the $r_M$
inconsistency as an additional systematic on the combined charge radius,
yielding a final combined result for the charge radius of $r_E=0.879(11)$~fm.
For the magnetic radius, we simply take the weighted average and apply half of
the difference between the Mainz and world results (0.034~fm) as an additional
uncertainty. For the charge radius, this additional uncertainty yields a 20\%
increase in the total uncertainty, while it is the dominant uncertainty in the
extraction of the magnetic radius.

\begin{table}[htbp]
\begin{center}
\begin{tabular}{l|l|l}
Source	& $r_E$	& $r_M$ \\
	& [fm]  & [fm]  \\ \hline

\textit{Published results} & & \\ \hline
$\mu$H~\cite{antognini13}		&~0.8409(4)	&~0.870(60) \\
$e$H~\cite{mohr12}			&~0.8758(77)	&~~~~~~- \\
Mainz A1~\cite{bernauer13,priv_distler}	&~0.8790(110)	&~0.777(19) \\
Zhan~\cite{zhan11}			&~0.8750(100)	&~0.867(20) \\
Sick~\cite{sick12, sick06}		&~0.8870(80)	&~0.855(35) \\
CODATA12 average~\cite{mohr12}		&~0.8775(51)	&~~~~~~-    \\ \hline
\textit{New updates} & & \\ \hline
Mainz updated				&~0.8750(150)	&~0.799(28) \\
world updated				&~0.8810(110)	&~0.867(20) \\
naive global average			&~0.8790(90)	&~0.844(16)\\
suggested global average		&~0.8790(110)	&~0.844(38) \\

\end{tabular}
\caption{Charge and magnetic radii from various published extractions and from
the averaging procedures described above. For ease of comparison, all charge
(magnetic) radii are quoted to 4 (3) significant digits.}
\label{tab:radii}
\end{center}
\end{table}

\section{Conclusions}\label{sec:summary}

We provide recommended radii and uncertainties based on electron scattering
measurements, taking into account the tension between different data sets.
Note that the previous CODATA evaluation~\cite{mohr12} included the extraction
of Ref.~\cite{bernauer10}, which did not include the hadronic TPE contribution,
and an earlier evaluation of world data~\cite{sick03} which had a
smaller data set and larger uncertainty than the extractions of~\cite{zhan11,
sick12}. While we include additional data compared to the previous evaluation,
our final uncertainty is noticeably larger as additional uncertainties have
been applied to account for more detailed evaluation of some uncertainties and
to account for tension between different extractions.

Because of this tension, we are forced to make a somewhat arbitrary choice in
evaluating the combined uncertainties, and have attempted to be somewhat
conservative in the uncertainties we choose. Further examination of possible
systematic uncertainties is presented in Refs.~\cite{sick12, lee15,
arrington_theseproc}, but at this point the origin of the discrepancy between
different data sets is not yet clear. We have laid out our assumptions and
reasoning for the additional uncertainties included but without an
understanding of the source of this discrepancy, it is not possible to come up
with a more rigorous way of evaluating the combined result and uncertainty.
We consider the approach presented above to be a reasonable combined average
of all e-p scattering data, and even with this significantly more conservative
approach, we still find a 3.5$\sigma$ discrepancy between e-p scattering
and muonic hydrogen measurements and a 5.7$\sigma$ discrepancy when combining
the electron scattering and atomic hydrogen transitions.

This work was supported in part by the U.S. Department of Energy, Office of
Science, Office of Nuclear Physics, under contract DE-AC-06CH11357

\bibliography{Arrington_Sick_Radius}

\end{document}